\documentclass[12pt]{article}

\usepackage{latexsym}
\usepackage{amsmath}
\usepackage{amssymb}
\usepackage{graphicx}
\usepackage{longtable}

\usepackage{bm}
\usepackage{epsfig}
\usepackage{citesort}

\begin{document}

\title{2+1 Flavor Polyakov--Nambu--Jona-Lasinio Model at Finite
Temperature and Nonzero Chemical Potential}

\author{{Wei-jie Fu$^{1}$, Zhao Zhang$^{1,2}$,
and Yu-xin Liu$^{1,2,3}$\thanks{corresponding author} }\\[3mm]
\normalsize{$^1$ Department of Physics and the State Key Laboratory
of Nuclear Physics }\\
\normalsize{and Nuclear Technology, Peking University, Beijing 100871, China}\\
\normalsize{$^2$ CCAST (World Laboratory), P.O. Box 8730, Beijing
100080, China} \\
\normalsize{$^3$ Center of Theoretical Nuclear Physics, National
Laboratory of Heavy Ion Accelerator,}\\ \normalsize{ Lanzhou 730000,
China}  }

\maketitle

%
%
%


\begin{abstract}
We extend the Polyakov-loop improved Nambu--Jona-Lasinio (PNJL)
model to 2+1 flavor case to study the chiral and deconfinement
transitions of strongly interacting matter at finite temperature and
nonzero chemical potential. The Polyakov-loop, the chiral
susceptibility of light quarks ($u$ and $d$) and the strange quark
number susceptibility as functions of temperature at zero chemical
potential are determined and compared with the recent results of
Lattice QCD simulations. We find that there is always an inflection
point in the curve of strange quark number susceptibility
accompanying the appearance of the deconfinement phase, which is
consistent with the result of Lattice QCD simulations. Predictions
for the case at nonzero chemical potential and finite temperature
are made as well.
%
%
We give the phase diagram in terms of the chemical potential and
temperature and find that the critical endpoint (CEP) moves down to
low temperature and finally disappears with the decrease of the
strength of the 't Hooft flavor-mixing interaction.
\end{abstract}



\section{Introduction}
\vspace{5pt}

QCD phase diagram and thermodynamics has been a subject of intense
investigation in recent years. Lattice QCD simulation is a
principal approach to explore the qualitative features of strongly
interacting matter and make quantitative prediction of its
properties. Over the years, this formulation has given us a wealth
of information about the phase diagram and thermodynamics at
finite temperature and limited chemical potential(see for example
Refs.~\cite{Boyd1996,Engels1999,Fodor2002,Fodor2003,Allton2002,Allton2003,Allton2005,Laermann2003,deForcrand2003,Kratochvila0456,delia1,delia2,ADGL05,AFHL05,Aoki2006a,Aoki2006b}.
In response to the Lattice QCD simulations, many phenomenological
approaches in terms of effective degrees of freedom have been
developed to give interpretation of the available Lattice data and
further to make prediction in the region of phase diagram that
can't be reached by the Lattice QCD.

A promising ansatz of this sort approach is the Polyakov-loop
improved Nambu--Jona-Lasinio model
(PNJL)~\cite{Meisinger9602,Pisarski2000,Fukushima2004,Ratti2006a},
which combines the two principal non-perturbative features of
low-energy QCD: confinement and spontaneous chiral symmetry
breaking in a simple formalism. The validity of PNJL model has
been tested in a series works by confronting the PNJL results with
the Lattice QCD
data~\cite{Ratti2006a,Ghosh2006,Ratti2006c,Ratti2006b,zhang2006}.
It has been reported that the 2 flavor ($u$ and $d$ quarks) PNJL
model can reproduce the result that the crossovers for
deconfinement phase transition and the chiral phase transition
almost coincide~\cite{Fukushima2004,Ratti2006a}. For finite
chemical potential, further investigations suggest that the
thermodynamics and susceptibilities obtained in PNJL model are
perfectly in agreement with the corresponding Lattice QCD
data~\cite{Ratti2006a,Ghosh2006,Ratti2006c}. Recently, the impact
of Polyakov-loop dynamics on the color superconductivity phase
transition and on the pion superfluidity phase transition within
PNJL have been explored in Refs.~\cite{Ratti2006b} and
\cite{zhang2006}, respectively.

Although the entanglement of the chiral and the Polyakov-loop
dynamics turns out to be indispensable to understand the nature of
QCD phase transitions and thermodynamical behavior, the
investigations based on this idea up to now are all performed  for
the two flavor case with small current quark mass so far. Whether
the synthesis of Polyakov-loop dynamics with the NJL model when
including strange quark works well is still unknown. Neverthelss,
the study of the critical temperature of the 2+1 flavor QCD phase
transition by the Lattice QCD simulations with physical masses in
the continuum limit have recently been
reported~\cite{Cheng2006,Aoki2006a}. Therefore it is interesting to
extend the 2 flavor PNJL model to the 2+1 flavor (i.e., including
not only $u$, $d$ quarks, but also $s$ quark) case to compare with
the recent results of the Lattice simulations. In addition, it is
worth investigating the 2+1 flavor QCD phase diagram at finite
baryon chemical potential in the NJL model with Polyakov-loop.

One should note that, in the absence of dynamical quarks, the
Polyakov-loop expectation value can be taken as an order parameter
to identify the color confinement and deconfinement in pure gauge
theories (see for example
Refs.~\cite{Meisinger9602,Pisarski2000,Fukushima2004,Ratti2006a}).
With the introduction of dynamical quarks with physical masses, the
exact $Z(N_{c})$ center symmetry for the gauge fields is lost and
the Polyakov-loop is no longer an exact order parameter for the
transition from the low temperature, confined phase, to the high
temperature, deconfined phase. However, with the distribution
function of the quark states, the Polyakov-loop expectation value
can still describe the evolution from color-singlet tri-quark state
to color-triplet single quark state. Moreover, the results of
Lattice QCD~\cite{Aoki2006a} show that the Polyakov-loop is also an
useful quantity to locate the crossover for deconfinement. Then, the
terminology ``confinement" we take here is the statistical
suppression of the color-triplet quark propagation, but not the
dynamical quark confinement. Similarly, the chiral symmetry is
explicitly broken by the nonvanishing current masses of quarks.
However the chiral condensate is also useful to indicate the chiral
crossover.

The purpose of this paper is to extend the 2 flavor PNJL model to
2+1 flavor case to explore the QCD thermodynamics and phase diagram
at finite temperature and nonzero chemical potential. The focuses
are concentrated on the validity of 2+1 flavor PNJL at finite
temperature in comparison with Lattice QCD data and its predictions
for the case at nonzero chemical potential and finite temperature.
The paper is organized as follows. In Section 2, we extend the PNJL
model to the case of 2+1 flavors. In Section 3, we discuss the phase
transition of strongly interacting matter at finite temperature and
zero chemical potential by analyzing the Polyakov-loop, the chiral
susceptibility of light quarks, the strange quark number
susceptibility and other related quantities as functions of
temperature at zero chemical potential and comparing the results
with those of Lattice QCD simulations. In Section 4, we make
predictions for the phase transitions in the case at nonzero
chemical potential and finite temperature and give a phase diagram
of the strongly interacting matter in terms of the temperature and
chemical potential. Finally, in Section 5, we give a summary and
conclusion.

\section{The 2+1 Flavor PNJL Model}

Following Ref. \cite{Ratti2006a}, we extend the 2 flavor
Polyakov-loop improved NJL model to include the strange quark and
the 2+1 flavor NJL model \cite{Kunihiro1989} with a Polyakov-loop
can then be given as
\begin{eqnarray}
\mathcal{L}_{PNJL}&=&\bar{\psi}\left(i\gamma_{\mu}D^{\mu}+\gamma_{0}
 \hat{\mu}-\hat{m}_{0}\right)\psi
 +G\sum_{a=0}^{8}\left[\left(\bar{\psi}\tau_{a}\psi\right)^{2}
 +\left(\bar{\psi}i\gamma_{5}\tau_{a}\psi\right)^{2}\right]\notag   \nonumber \\
&&-K\left[\textrm{det}_{f}\left(\bar{\psi}\left(1+\gamma_{5}\right)\psi\right)
 +\textrm{det}_{f}\left(\bar{\psi}\left(1-\gamma_{5}\right)\psi\right)\right]
 -\mathcal{U}\left(\Phi,\Phi^{*} \, ,T\right),\label{lagragian}
\end{eqnarray}
where $\psi=(\psi_{u},\psi_{d},\psi_{s})^{T}$ is the three-flavor
quark field,
\begin{equation}
D^{\mu}=\partial^{\mu}-iA^{\mu}\quad\textrm{with}\quad
A^{\mu}=\delta^{u}_{0}A^{0}\quad\textrm{,}\quad
A^{0}=g\mathcal{A}^{0}_{a}\frac{\lambda_{a}}{2}=-iA_4.
\end{equation}
$\lambda_{a}$ are the Gell-Mann matrices in color space.
$\hat{m}_{0}=\textrm{diag}(m_{u},m_{d},m_{s})$ is the three-flavor
current quark mass matrix. Throughout this work, we take
$m_{u}=m_{d}\equiv m_{l}$, assuming the isospin symmetry is
reserved on the Lagrangian level, whereas $m_{s}$ is usually
different from $m_{l}$, thus the $SU(3)_f$ symmetry is explicitly
broken. The quark chemical potential matrix $\hat{\mu}$ is chosen
to proportional to the unit matrix in our work for simplicity,
namely, these three flavor quarks have identical chemical
potential $\mu$. In the above PNJL Lagrangian, the four-point
interaction term with an effective coupling strength $G$ is
$U(3)_{L}\otimes U(3)_{R}$ symmetric, where
$\tau_{0}=\sqrt{\frac{2}{3}}\textit{1}_{f}$ and $\tau_{a}$,
$a=1,\ldots,8$ are the eight Gell-Mann matrices in flavor space.
The flavor-mixing term with coupling strength $K$ is a determinant
in flavor space, which corresponds to the 't Hooft interaction. It
breaks the $U_{A}(1)$ symmetry and leaves $SU(3)_{L}\otimes
SU(3)_{R}$ unbroken. $\mathcal{U}\left(\Phi,\Phi^{*},T\right)$ is
the Polyakov-loop effective potential.

The above description shows apparently that the degrees of freedom
for temporal gauge field expressed in a spatially homogeneous
background field is explicitly included in the PNJL model,
compared with the conventional NJL model. The Polyakov-loop
dynamics represented by this background field is controlled by the
Polyakov-loop effective potential $\mathcal{U}(\Phi,\Phi^{*},T)$
and the quarks which couple to the Polyakov-loop. This effective
potential can be expressed in terms of the Polyakov-loop
expectation value (or, in other word, the traced Polyakov-loop)
$\Phi=(\mathrm{Tr}_{c}L)/N_{c}$ and its conjugate
$\Phi^{*}=(\mathrm{Tr}_{c}L^{\dag})/N_{c}$ with the Polyakov-loop
$L$ being a matrix in color space given explicitly
by~\cite{Ratti2006a}
\begin{equation}
L\left(\vec{x}\right)=\mathcal{P}\exp\left[i\int_{0}^{\beta}d\tau\,
A_{4}\left(\vec{x},\tau\right)\right]
                     =\exp\left[i \beta A_{4} \right]\, ,
\end{equation}
with  $\beta=1/T$ being the inverse of temperature and
$A_{4}=iA^{0}$. In the so-called Polyakov gauge, the Polyakov-loop
matrix can be given as a diagonal representation
\cite{Fukushima2004}. The coupling between Polyakov-loop and quarks
is uniquely determined by the covariant derivative $D_{\mu}$ in the
PNJL Lagrangian in Eq.~\eqref{lagragian}. The trace of the
Polyakov-loop, $\Phi$ and its conjugate, $\Phi^{*}$, can be handled
with classical field variables in the PNJL.

Temperature dependent effective potential
$\mathcal{U}(\Phi,\Phi^{*},T)$ is taken to reproduce the
thermodynamical behavior of the Polyakov-loop for the pure gauge
case in accordance with Lattice QCD predictions, and it has the
$Z(3)$ center symmetry like the pure gauge QCD Lagrangian. This
$Z(3)$ center symmetry is spontaneously broken when temperature is
above some critical point ($T_{0}\simeq 270\,\mathrm{MeV}$ in pure
gauge QCD \cite{Ratti2006a}) and the traced Polyakov-loop develops a
finite value. In the absence of quarks, $\Phi=\Phi^{*}$ and the
Polyakov-loop serves as an order parameter for the deconfinement
exactly.

In previous works, two possible forms for the Polyakov-loop
effective potential have been well developed. One is a polynomial in
$\Phi$ and $\Phi^{*}$ \cite{Ratti2006a}, which is denoted by
$\mathcal{U}_{pol}(\Phi,\Phi^{*},T)$ in this work and another is an
improved effective potential in which the higher order polynomial
terms in $\Phi$ and $\Phi^{*}$ are replaced by a
logarithm~\cite{Ratti2006c,Ratti2006b}. We denote this improved
effective potential by $\mathcal{U}_{imp}(\Phi,\Phi^{*},T)$. Both
effective potentials are taken in the present work to investigate
whether our results depend on the details of the Polyakov-loop
effective potential. These two effective potentials have the
following forms
\begin{equation}
\frac{\mathcal{U}_{pol}\left(\Phi,\Phi^{*},T\right)}{T^{4}} =
-\frac{b_{2}(T)}{2}\Phi^{*}\Phi -\frac{b_{3}}{6}
(\Phi^{3}+{\Phi^{*}}^{3})+\frac{b_{4}}{4}(\Phi^{*}\Phi)^{2} \, ,
\end{equation}
with
\begin{equation}
b_{2}(T)=a_{0}+a_{1}\left(\frac{T_{0}}{T}\right)+a_{2}
{\left(\frac{T_{0}}{T}\right)}^{2}
+a_{3}{\left(\frac{T_{0}}{T}\right)}^{3},
\end{equation}
and
\begin{eqnarray}
\frac{\mathcal{U}_{imp}\left(\Phi,\Phi^{*},T\right)}{T^{4}} & =
&-\frac{1}{2}a(T)\Phi^{*}\Phi 
+b(T)\ln\left[1-6\Phi^{*}\Phi+4({\Phi^{*}}^{3}+\Phi^{3})
-3(\Phi^{*}\Phi)^{2}\right]  \, ,
\end{eqnarray}
with
\begin{equation}
a(T)=a_{0}+a_{1}\left(\frac{T_{0}}{T}\right)
     +a_{2}{\left(\frac{T_{0}}{T}\right)}^{2},\quad
b(T)=b_{3}{\left(\frac{T_{0}}{T}\right)}^{3}.
\end{equation}
A precise fit of the parameter $a_{i}$ and $b_{i}$ in these two
effective potentials has recently been performed to reproduce the
Lattice QCD data for pure gauge QCD thermodynamics and the
behavior of the Polyakov-loop as a function of temperature in
Refs.~\cite{Ratti2006a} and \cite{Ratti2006b}, respectively. In
these two works, $T_{0}=270\,\mathrm{MeV}$ is chosen to be the
critical temperature for the deconfinement to take place in the
pure gauge QCD.

After performing the mean field approximation for the Lagrangian of
the three-flavor PNJL in Eq.~\eqref{lagragian}, we obtain the
thermodynamical potential density as
\begin{eqnarray}
\Omega&=&\mathcal{U}\left(\Phi,\Phi^{*},T\right)+2G\left({\phi_{u}}^{2}
+{\phi_{d}}^{2}+{\phi_{s}}^{2}\right)-4K\phi_{u}\,\phi_{d}\,\phi_{s} \nonumber \\
&&-2\int\frac{\mathrm{d}^{3}p}{\left(2\pi\right)^{3}}
\left\{3\left(E_{u}+E_{d}+E_{s}\right)
\theta\left(\Lambda^{2}-p^{2}\right)\right. \nonumber \\
&&{}+T\ln\left[1+3\Phi e^{-\left(E_{u}-\mu\right)/T}+3\Phi^{*}
e^{-2\left(E_{u}-\mu\right)/T}+e^{-3\left(E_{u}-\mu\right)/T}\right]\nonumber \\
&&{}+T\ln\left[1+3\Phi^{*} e^{-\left(E_{u}+\mu\right)/T}+3\Phi
e^{-2\left(E_{u}+\mu\right)/T}+e^{-3\left(E_{u}+\mu\right)/T}\right] \nonumber \\
&&{}+T\ln\left[1+3\Phi e^{-\left(E_{d}-\mu\right)/T}+3\Phi^{*}
e^{-2\left(E_{d}-\mu\right)/T}+e^{-3\left(E_{d}-\mu\right)/T}\right] \nonumber \\
&&{}+T\ln\left[1+3\Phi^{*} e^{-\left(E_{d}+\mu\right)/T}+3\Phi
e^{-2\left(E_{d}+\mu\right)/T}+e^{-3\left(E_{d}+\mu\right)/T}\right] \nonumber  \\
&&{}+T\ln\left[1+3\Phi e^{-\left(E_{s}-\mu\right)/T}+3\Phi^{*}
e^{-2\left(E_{s}-\mu\right)/T}+e^{-3\left(E_{s}-\mu\right)/T}\right] \nonumber  \\
&&\left.+T\ln\left[1+3\Phi^{*} e^{-\left(E_{s}+\mu\right)/T}+3\Phi
e^{-2\left(E_{s}+\mu\right)/T}+e^{-3\left(E_{s}+\mu\right)/T}\right]\right\},\label{thermopotential}
\end{eqnarray}
where $\mu$ is the chemical potential, $T$ is the temperature,
$\phi_{i}$ is the chiral condensate of quarks with flavor $i$, and
$E_{i}=\sqrt{p^{2}+M_{i}^{2}}$ is its corresponding quasiparticle
energy, with constituent masses for the quark of flavor $i$
\begin{equation}
M_{i}=m_{i}-4G\phi_{i}+2K\phi_{j}\,\phi_{k}.\label{constituentmass}
\end{equation}
 As mentioned above, the breaking of the isospin symmetry is
neglected throughout this work. We have thus $\phi_{u}=\phi_{d}
\,\equiv \phi_{l}$ in the absence of isospin chemical potential.

Minimizing the thermodynamical potential in
Eq.~\eqref{thermopotential} with respective to $\phi_{l}$,
$\phi_{s}$, $\Phi$, and $\Phi^{*}$, we obtain a set of equations of
motion
\begin{equation}
\frac{\partial\Omega}{\partial\phi_{l}}=0, \quad
\frac{\partial\Omega}{\partial\phi_{s}}=0, \quad
\frac{\partial\Omega}{\partial\Phi}=0, \quad
\frac{\partial\Omega}{\partial\Phi^{*}}=0.\label{motion_equations}
\end{equation}
This set of equations can be solved for the fields as functions of
temperature $T$ and chemical potential $\mu$. It has been shown in
Ref.~\cite{Ratti2006a} that, in 2 flavor PNJL model, there is
$\Phi=\Phi^{*}$ when the quark chemical potential is vanishing
($\mu=0$). While for $\mu\neq0$, $\Phi$ and $\Phi^{*}$ have
different values.

In the NJL sector of this model, five parameters need to be
determined. In our present work we adopt the parameter set in
Ref.~\cite{Rehberg1996}, $m_{l}=5.5\;\mathrm{MeV}$,
$m_{s}=140.7\;\mathrm{MeV}$, $G\Lambda^{2}=1.835$,
$K\Lambda^{5}=12.36$ and $\Lambda=602.3\;\mathrm{MeV}$, which is
fixed by fitting $m_{\pi}=135.0\;\mathrm{MeV}$,
$m_{K}=497.7\;\mathrm{MeV}$,
$m_{\eta^{\prime}}=957.8\;\mathrm{MeV}$ and
$f_{\pi}=92.4\;\mathrm{MeV}$.
%

\section{Phase Transition in the Case of $\mu=0$ and $T\neq0$}

It has been strongly suggested by current Lattice QCD simulations
that the transition from low temperature hadronic phase to high
temperature quark-gluon-plasma (QGP) phase at vanishing quark
chemical potential is a continuous, non-singular but rapid
crossover \cite{Aoki2006b,Karsch2002}. It has also been
demonstrated that because of the non-singularity of the crossover,
different observables lead to various values of transition
temperature $(T_{c})$ in the 2+1 flavor QCD with physical masses
both for the light quarks $m_{l}$ and for the strange quark
$m_{s}$ even in the continuum and thermodynamical
limit~\cite{Aoki2006a}. In order to determine the critical
temperature $T_{c}$ at vanishing quark chemical potential in the
$2+1$ flavor PNJL model, we consider three quantities, which were
used to locate the transition point in the Lattice QCD simulations
in Ref.~\cite{Aoki2006a}, the traced Polyakov-loop (i.e., the
Polyakov-loop expectation value. In the following we just denote
it as Polyakov-loop for simplicity), the light quark chiral
susceptibility and the strange quark number susceptibility.

The chiral susceptibility of the light quark is defined as
\begin{equation}
\chi_{l}=-\frac{\partial^{2}\Omega}{\partial\, {m_{l}}^{2}}.
\end{equation}
In order to obtain a dimensionless quantity and renormalize the
divergence of the thermodynamical potential $\Omega$ in Lattice
QCD simulations, this quantity was normalized to the following one
\cite{Aoki2006a}:
\begin{equation}
\frac{{m_{l}}^{2}\Delta\chi_{l}}{T^{4}}=
\frac{{m_{l}}^{2}}{T^{4}}\left[\chi_{l}\left(T\right)
-\chi_{l}\left(0 \right)\right].
\end{equation}
In the PNJL model we have
\begin{eqnarray}
\frac{{m_{l}}^{2}\chi_{l}}{T^{4}}&=&\frac{6{m_{l}}^{2}}{T^{4}}
\int\frac{\mathrm{d}^{3}p}{\left(2\pi\right)^{3}}\theta\left(\Lambda^{2}-p^{2}\right)
\left\{\frac{p^{2}}{{E_{l}}^{3}}\left[1-2f\left(E_{l},T,\Phi\right)\right]\right. \nonumber \\
&&\left.{}+\frac{2}{T}\left(\frac{M_{l}}{E_{l}}\right)^{2}\left[\frac{\Phi
e^{-E_{l}/T}+4\Phi e^{-2E_{l}/T}
+3e^{-3E_{l}/T}}{A\left(E_{l},T,\Phi\right)}
-3f^{2}\left(E_{l},T,\Phi\right)\right]\right\},
\end{eqnarray}
with $A\left(x,T,\Phi\right)$ and $f\left(x,T,\Phi\right)$ being
defined as
\begin{equation}
A(x,T,\Phi)=1+3\Phi e^{-x/T}+3\Phi e^{-2x/T}+e^{-3x/T}\label{A} \, ,
\end{equation}
and
\begin{equation}
f\left(x,T,\Phi\right)=\frac{\Phi e^{-x/T}+2\Phi
e^{-2x/T}+e^{-3x/T}}{1+3\Phi e^{-x/T}+3\Phi e^{-2x/T}+e^{-3x/T}} \,
,  \label{DistFunct}
\end{equation}
respectively. Here, we do not distinguish $\Phi$ and $\Phi^{*}$,
because they have the same value at $\mu=0$.

The strange quark number susceptibility is defined as
\cite{Bernard2005}
\begin{equation}
\frac{\chi_{s}}{T^{2}}=-\frac{1}{T^{2}}\frac{\partial^{2}\Omega}
{\partial\:{\mu_{s}}^{2}}\Big|_{\mu_{s}=0} \, ,
\end{equation}
and in the PNJL model given explicitly by
\begin{equation}
\frac{\chi_{s}}{T^{2}} = \frac{12}{T^{3}}\int\frac{\mathrm{d}^{3}p}
{\left(2\pi\right)^{3}} \left[\frac{\Phi e^{-E_{s}/T}+4\Phi
e^{-2E_{s}/T}+3e^{-3E_{s}/T}}{A\left(E_{s},T,\Phi\right)}
-3f^{2}\left(E_{s},T,\Phi\right)\right].
\end{equation}

\begin{figure}[ht]
\begin{center}
\includegraphics[scale=0.8]{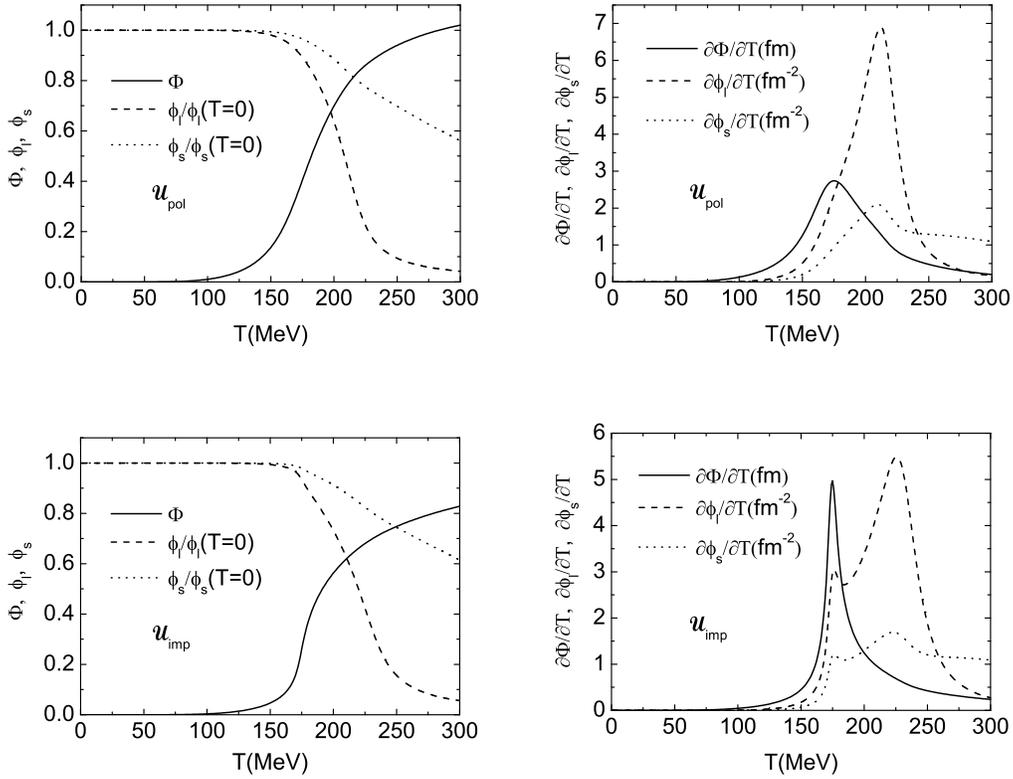}
\end{center}
\caption{Left panel: calculated Polyakov-loop $\Phi$, scaled chiral
condensates for light quarks $\phi_{l}$ and for strange quarks
$\phi_{s}$ as functions of temperature at vanishing quark chemical
potential in the case of considering the polynomial Polyakov-loop
effective potential $\mathcal{U}_{pol}$, the improved Polyakov-loop
effective potential $\mathcal{U}_{imp}$, respectively. Right panel:
calculated temperature dependence of the derivatives
$\partial\Phi/\partial T$, $\partial\phi_{l}/\partial T$ and
$\partial\phi_{s}/\partial T$ with the two kinds of Polyakov-loop
potentials. } \label{f1}
\end{figure}

The calculated results of the temperature dependence of the
Polyakov-loop, the light quark chiral condensate and the strange
quark chiral condensate at zero quark chemical potential with the
usual polynomial Polyakov-loop effective potential
$\mathcal{U}_{pol}$ and the improved Polyakov-loop effective
potential $\mathcal{U}_{imp}$ are shown in the left panel of
Fig.~\ref{f1}. As we can see, when quarks (including the light and
the strange) have physical current masses, the chiral phase
evolution is not a true phase transition but a continuous
crossover. At the same time, the coupling of the Polyakov-loop to
quarks turns the first-order deconfinement transition  for the
pure-gauge QCD into a crossover as the same as that in the two
flavor PNJL case~\cite{Ratti2006a}. To determine the
pseudo-transition temperature, we have also calculated the
derivatives of the Polyakov-loop and the chiral condensates with
respect to temperature. The obtained results with the two kinds of
Polyakov-loop effective potentials are illustrated in the right
panel of Fig.~\ref{f1}. In order to confront our results to those
in the Lattice QCD simulations with physical masses for the 2+1
flavor QCD~\cite{Aoki2006a}, we have rescaled the parameter
$T_{0}$ in $\mathcal{U}_{pol}$ from $270$ to $200$~MeV and in
$\mathcal{U}_{imp}$ from $270$ to $215$~MeV along the way taken in
Ref.\cite{Ratti2006a}. After such a rescaling, we obersve that
both the Polyakov-loop effective potentials give the same
pseudo-deconfinement transition temperature $175$~MeV (see right
panel of Fig.~\ref{f1}). This value is consistent with the Lattice
QCD result $T_{c}(P)=176(3)(4)\:\mathrm{MeV}$~\cite{Aoki2006a},
where the numbers in the parenthesis indicate the errors.

After rescaling the $T_{0}$ in the effective potential, we can
determine the pseudo-critical point $T_{c}$ for the chiral phase
transition with other quantities. From the appearance of the peaks
of $\partial\phi_{l}/\partial T$ and $\partial\phi_{s}/\partial T$
shown in the right panel of Fig.~\ref{f1}, one can infer that the
pseudo-critical temperature of the chiral phase transition for light
quarks $T_{c}(l)$ are almost the same as that of strange quark
$T_{c}(s)$. Such a similarity is quite natural due to the
flavor-mixing interaction in the Lagrangian shown in
Eq.~\eqref{lagragian}. In more details, the critical temperatures
obtained from the polynomial effective potential are
$T_{c}(l)=212\:\mathrm{MeV}$ and $T_{c}(s)=210\:\mathrm{MeV}$, while
the corresponding temperatures from the improved effective potential
are $226$~MeV and $223$~MeV. These values are all relatively larger
than the recent Lattice QCD results, which are
$T_{c}(\chi_{l})=192(7)(4)\:\mathrm{MeV}$ in Ref.~\cite{Cheng2006},
$151(3)(3)$~MeV in Ref.~\cite{Aoki2006a} and $169(12)(4)$~MeV in
Ref.~\cite{Bernard2005}. This feature is the same as the standard
NJL model which also has relatively larger chiral transition
temperature~\cite{Buballa2005}.

Comparing the upper panel with the lower panel of Fig.~\ref{f1},
one can find that the Polyakov-loop corresponding to the improved
effective potential changes more rapidly near the $T_{c}(P)$ than
that relevant to the polynomial effective potential. The peak of
$\partial \Phi/\partial T$ in the lower-right panel of
Fig.~\ref{f1} is much narrower and higher, which induces two
little bumps in the curves of $\partial\phi_{l}/\partial T$ and
$\partial\phi_{s}/\partial T$ at $T_{c}(P)$, while there is no
such bump in the curves in the upper-right panel of Fig.~\ref{f1}
for the polynomial effective potential. These phenomena indicate
that, for the improved effective potential, the deconfinement
transition has an obvious influence on the chiral condensate.

One may have a question about why the peak of the temperature
derivative of the Polyakov-loop $\Phi$ obtained in the improved
effective potential is much narrower than that given by the
polynomial effective potential. To explore the underlying physics,
we recall that, once the dynamical quarks with physical masses are
introduced, the Z(3) center symmetry is explicitly broken, so that
deconfinement phase transition changes from a first order one in
pure gauge theories to a crossover. It is then natural that the
temperature derivative of the Polyakov-loop smears to a peak with
finite height and finite width. The calculated results shown in
Fig.~\ref{f1} manifest that both the polynomial effective potential
and the improved effective potential give definitely the temperature
derivative of the Polyakov-loop peaks with finite width and finite
height. It indicates that both the effective potentials represent
the dynamics correctly in some sense. As for that the width given by
the improved effective potential is narrower, we should note that
recent Lattice QCD results show that the normalized pressure for
full QCD with dynamical fermions looks the same as that in the pure
gauge theories~\cite{Karsch2002}, which indicates that it may be the
dynamics of gluons that drives the QCD phase transition, but not
that of the fermions~\cite{Rischke2004}. And the crossover of the
phase evolution is a quite rapid one~\cite{Aoki2006b,Karsch2002}.
The improved effective potential replaces the higher order terms in
$\Phi$ and $\Phi^{*}$ in the polynomial effective potential by the
logarithm of the Jacobi determinant which results from integrating
out the six non-diagonal Lie algebra directions while keeping the
two diagonal ones to represent
$\Phi$~\cite{Fukushima2004,Ratti2006c,Ratti2006b}. It means that,
the improved effective potential describes the dynamics of the
Polyakov-loop (i.e., the gluon dynamics) more appropriately and
correlates to the effect of the dynamical quarks more slightly. The
narrower width of the deconfinement crossover with the improved
effective potential can thus be attributed to that the improved
Polyakov-loop effective potential represents better the gluon
dynamics.

\begin{figure}[!htb]
\centering
\includegraphics[scale=1.0]{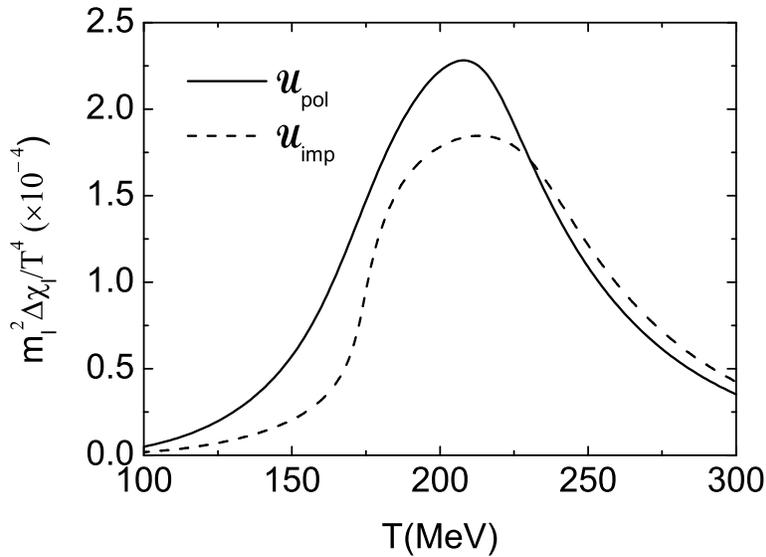}          
\caption{Calculated results of the renormalized chiral
susceptibility ${m_{l}}^{2}\Delta\chi_{l}/T^{4}$ as a function of
temperature at zero quark chemical potential for the two
Polyakov-loop effective potentials.} \label{f2}
\end{figure}

The calculated results of the temperature dependence of the chiral
susceptibility of light quarks are shown in Fig.~\ref{f2}. The
pseudo-critical point $T_{c}(\chi_{l})$ determined from the peak
of the chiral susceptibility is $208\:\mathrm{MeV}$ for
$\mathcal{U}_{pol}$ and $213$~MeV for $\mathcal{U}_{imp}$. These
values do not exactly coincide with the results  of $T_{c}(l)$s
shown in Fig.~\ref{f1}, which reflects the nature of the crossover
of the phases. However, the discrepancies of the pseudo-critical
points from different susceptibilities are within $10$~MeV.

\begin{figure}[ht]
\centering
\includegraphics[scale=0.9]{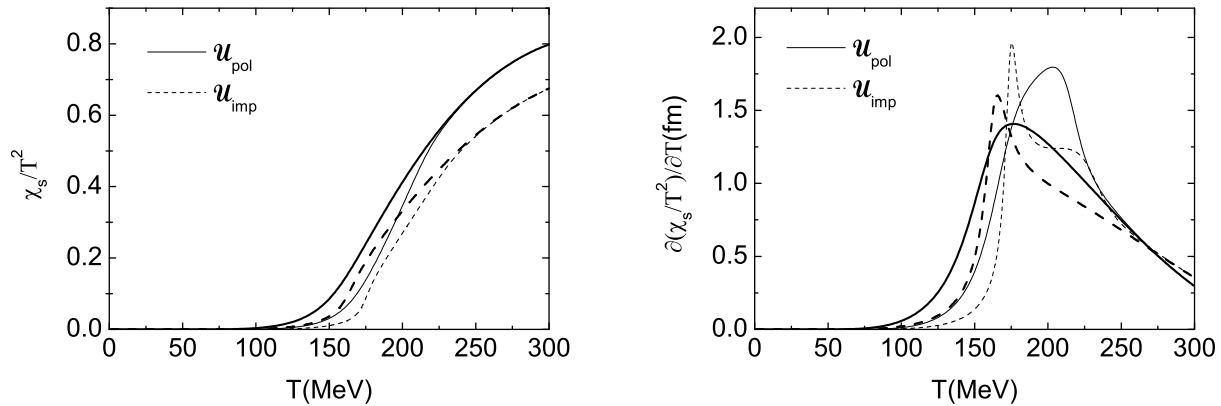}           
\caption{Left panel: calculated results of the scaled strange quark
number susceptibility $\chi_{s}/T^{2}$ as a function of temperature
(thinner curves are for the case with $K\Lambda^{5}=12.36$ and
thicker ones for $K\Lambda^{5}=0$). Right panel: plots of the
calculated $\partial(\chi_{s}/T^{2})/\partial T$. } \label{f3}
\end{figure}

Fig.~\ref{f3} shows the calculated temperature dependence of the
strange quark number susceptibility. First of all, we discuss the
case including the flavor-mixing effects with $K\Lambda^{5}=12.36$
in the Lagrangian in Eq.~\eqref{lagragian}. From the calculated
results shown as the thinner curves in Fig.~\ref{f3}, one can notice
that the $\chi_{s}/T^{2}$ increases monotonically with temperature.
Ref.~\cite{Aoki2006a} has shown that the pseudo-critical point
$T_{c}(\chi_{s})$ can be defined as the inflection point of a
susceptibility curve, viz. the peak of the curve of
$\partial(\chi_{s}/T^{2})/\partial T$. Our calculated results
illustrated in the right panel of Fig.~\ref{f3} manifest that the
location of this pseudo-critical point is affected by both the
deconfinement and the chiral crossovers. Because of the common
influences by these two crossovers, we find
$T_{c}(\chi_{s})=203\:\mathrm{MeV}$ for the polynomial potential,
which is in the between of the deconfinement pseudo-critical point
$T_{c}(P)=175\:\mathrm{MeV}$ and the chiral critical point
$T_{c}(l)=212\:\mathrm{MeV}$, $T_{c}(s)=210\:\mathrm{MeV}$. This
phenomenon is more obvious in the result of the improved effective
potential, where a sharp peak appears at the deconfinement
transition point $T_{c}(P)=175\,\mathrm{MeV}$ on the curve of
$\partial(\chi_{s}/T^{2})/\partial T$. It manifests evidently
$T_{c}(\chi_{s}) = T_{c}(P)$. In addition, the peak is followed by a
little platform coming from the contribution of the chiral crossover
for light quarks due to the flavor-mixing effects. This phenomenon
can be understood as the follows. Being not color-singlets, there
are almost no free strange quarks before the deconfinement crossover
takes place. As the temperature increases, the crossover for the
deconfinement occurs and the strange quark number susceptibility
develops to finite values abruptly. In general case, the
quasiparticle energy of strange quarks becomes smaller due to the
partial reduction of the condensate of strange quarks which is along
with the light quarks chiral crossover because of the flavor-mixing
interactions. It leads it easier to excite quarks, and in turn to
increase the value of the quark number susceptibility. However,
since the improved effective potential represents the gluon dynamics
better and the deconfinement crossover governed by the gluon
dynamics occurs more abruptly as mentioned above, the increasing
rate of the strange quark number susceptibility with respect to
temperature induced by the deconfinement crossover dominates over
that induced by the reduction of the quark condensate at the
deconfinement pseudo-critical temperature. As a consequence, the
pseudo-critical temperature given by the strange quark number
susceptibility coincides with that for the deconfinement crossover.
As the temperature increases further, the reduction of the
quasiparticle energy of the quark begins to play the role, it
generates then a platform in the curves of
$\partial(\chi_{s}/T^{2})/\partial T$. In order to study the
influence of the deconfinement crossover on the $\chi_{s}$ further
we turn off the flavor-mixing interaction by setting
$K\Lambda^{5}=0$. The obtained results are illustrated in the
thicker curves in Fig.~\ref{f3}. We find again that $\chi_{s}/T^{2}$
gradually develops a finite value with the appearance of the
deconfinement phase. Since the $\chi_{s}/T^{2}$ is no longer
influenced by the light quark chiral crossover any more, the
difference between the pseudo-critical temperatures given by the two
effective potentials is much smaller than that with the
flavor-mixing being taken into account. Furthermore, the curve of
$\partial(\chi_{s}/T^{2})/\partial T$ for the polynomial potential
moves down to lower temperature obviously and there are not any
platforms on the curve of $\partial(\chi_{s}/T^{2})/\partial T$ for
the improved effective potential. Therefore, we could qualitatively
understand the results of Lattice QCD simulations in
Ref.~\cite{Aoki2006a} that the pseudo-critical point determined from
the strange quark number susceptibility is quite close to that
determined from the Polyakov loop. Nevertheless, it is not obvious
that the quark number susceptibility has any relations with the
chiral crossover for light quarks, which maybe indicates that the
flavor-mixing interactions are relatively weak at high temperature.

\begin{figure}[ht]
\centering
\includegraphics[scale=0.9]{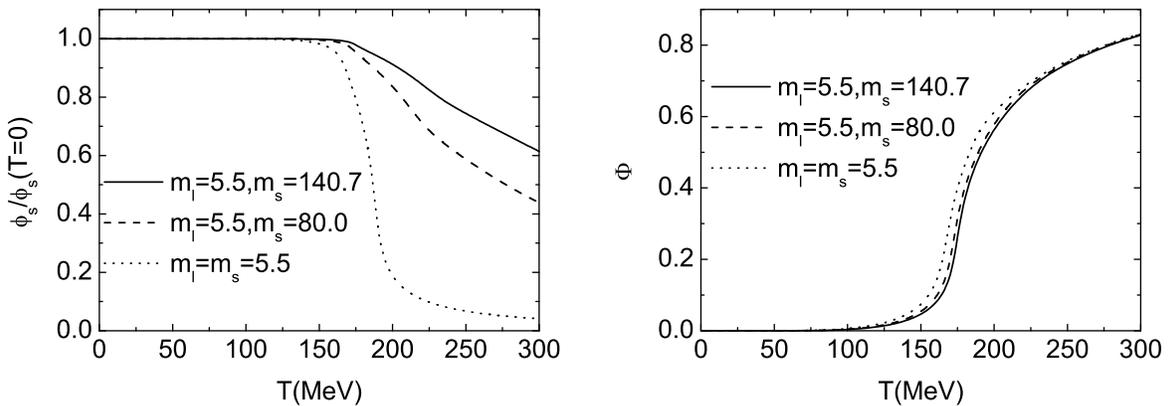}          
\caption{Calculated results of the scaled strange quark chiral
condensate as a function of temperature with different values of
the current mass of strange quark for the improved effective
potential (left panel) and the temperature dependence of the
corresponding Polyakov-loop $\Phi$ at several strange quark masses
(right panel). } \label{f4}
\end{figure}

In order to investigate the influences of the current mass of
strange quark on the properties of chiral and deconfinement
crossover in the case of including the flavor-mixing effects, we
depict the scaled strange quark condensate and the Polyakov-loop
as functions of temperature with different values of strange quark
current mass in Fig.~\ref{f4}, using the improved effective
potential $\mathcal{U}_{imp}$. Obviously, one can find from the
figure that the chiral crossover is quite sensitive to the
variation of the current mass of strange quark. The chiral
transition temperature $T_{c}(s)$ defined above decreases from
about 225 MeV to 188 MeV for the improved effective potential
$\mathcal{U}_{imp}$ when the value of $m_{s}$ is reduced from
$140.7\:\mathrm{MeV}$ to $5.5\:\mathrm{MeV}$. The same thing
occurs for the polynomial effective potential $\mathcal{U}_{pol}$.
Furthermore, the region of crossover for the chiral phase
evolution takes place in a more narrow range of temperature.
Nevertheless, as for the deconfinement transition, differences
between the physical case with $m_{s}=140.7\:\mathrm{MeV}$ and the
hypothetical case with $m_{s}=5.5\:\mathrm{MeV}$ are very small.
Consequently, the difference between the chiral pseudo-critical
point $T_{c}(s)$ and the deconfinement pseudo-critical point
$T_{c}(P)$ becomes smaller when the current mass of the strange
quark decreases.

\begin{figure}[!htb]
\centering
\includegraphics[scale=1.0]{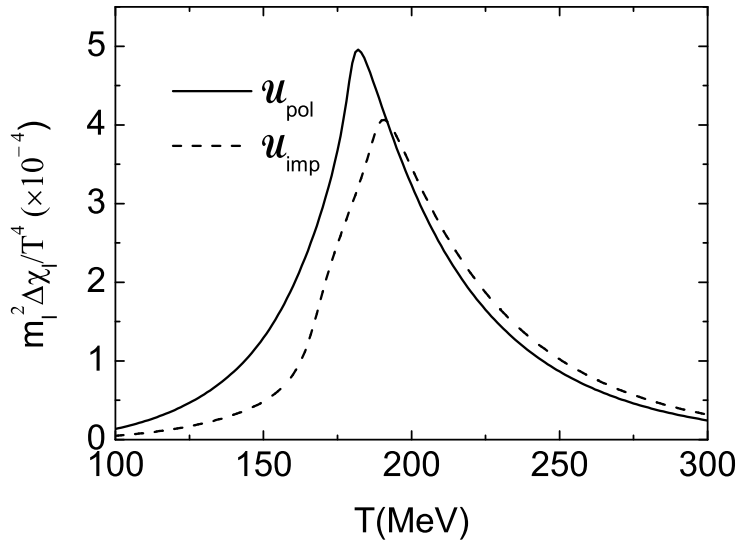}              
\caption{The same as Fig.~\ref{f2} but with
$m_{s}=m_{l}=5.5\:\mathrm{MeV}$.} \label{f5}
\end{figure}

Fig.~\ref{f5} shows the renormalized  chiral susceptibility
${m_{l}}^{2}\Delta\chi_{l}/T^{4}$ as a function of temperature for
the case of $m_{s}=m_{l}=5.5\:\mathrm{MeV}$, viz. neglecting the
flavor difference. As one expects, the curves in this figure are
much narrower and sharper than those in Fig.~\ref{f2}. The chiral
pseudo-critical temperature $T_{c}(\chi_{l})$ is $182$~MeV for the
$\mathcal{U}_{pol}$ and $191$~MeV for the $\mathcal{U}_{imp}$, which
are 26 MeV and 22 MeV smaller than their corresponding values for
the case with physical strange quark mass, respectively.

\begin{figure}[ht]
\centering
\includegraphics[scale=0.9]{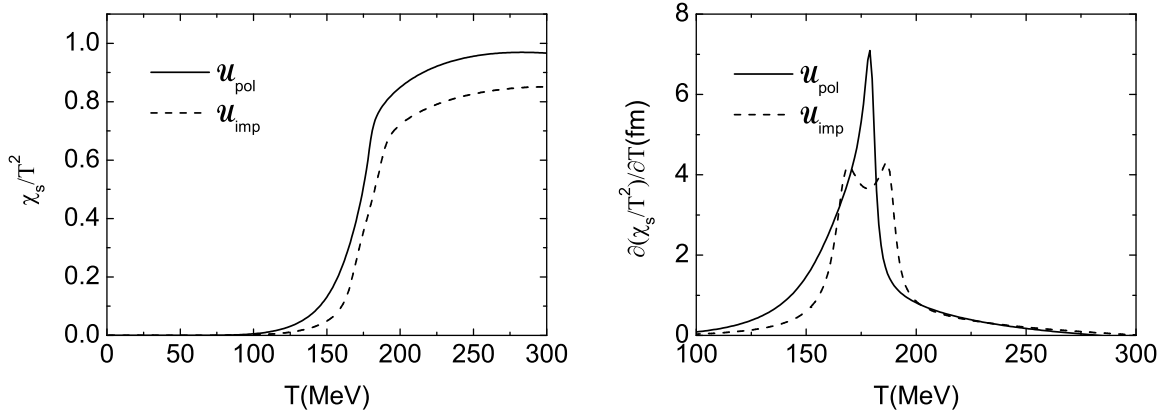}             
\caption{The same as Fig. \ref{f3} but with
$m_{s}=m_{l}=5.5\,\mathrm{MeV}$ and $K\Lambda^{5}=12.36$.}
\label{f6}
\end{figure}

\begin{figure}[!htb]
\centering
\includegraphics[scale=1.0,angle=0]{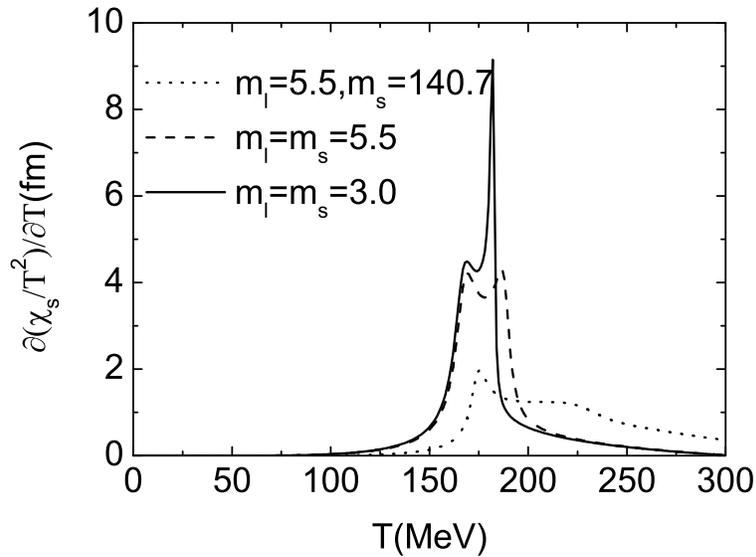}          
\caption{Calculated derivatives of the strange quark number
susceptibility versus the temperature in the cases:
$m_{l}=5.5\,\mathrm{MeV}$, $m_{s}=140.7\,\mathrm{MeV}$;
$m_{l}=m_{s}=5.5 \,\mathrm{MeV}$; and
$m_{l}=m_{s}=3.0\,\mathrm{MeV}$.} \label{f7}
\end{figure}

We have also calculated the temperature dependence of the strange
quark number susceptibility $\chi_{s}/T^{2}$ with
$m_{s}=m_{l}=5.5\:\mathrm{MeV}$. The obtained results are shown in
Fig.~\ref{f6}. From the figure we can recognize not surprisingly,
$T_{c}(\chi_{s})=179\,\mathrm{MeV}$ for the polynomial effective
potential, which is consistent with the chiral pseudo-critical
temperature $T_{c}(\chi_{l})=182\,\mathrm{MeV}$ and the
deconfinement critical temperature $T_{c}(P)=178\,\mathrm{MeV}$. As
for the improved effective potential, there are two peaks located at
$169$~MeV and $186$~MeV in the curve of the temperature derivative.
We have then the pseudo-critical temperature in the case of
$m_{s}=m_{l}=5.5\:\mathrm{MeV}$,
$T_{c}(\chi_{s})=169\,\mathrm{MeV}$, $186\,\mathrm{MeV}$, which
correspond to the $T_{c}(P)=168\,\mathrm{MeV}$,
$T_{c}(\chi_{l})=188\,\mathrm{MeV}$, respectively. Fig.~\ref{f7}
shows the temperature derivative of the strange quark number
susceptibility for the improved effective potential with several
values of current mass of strange quarks. One can observe from the
figure that the peak related to the chiral critical point grows
drastically as the current mass of strange quarks decreases from
$140.7$~MeV to $3.0$~MeV.

One should note that in order to confront our results to those in
Lattice QCD simulations in~\cite{Aoki2006a}, we have rescaled the
parameter $T_{0}$ in $\mathcal{U}_{pol}$ from $270$ to $200$~MeV and
in $\mathcal{U}_{imp}$ from $270$ to $215$~MeV to fix the
pseudo-deconfinement transition temperature
$T_{c}(P)=175\:\mathrm{MeV}$. However, recent Lattice QCD
simulations~\cite{Cheng2006,Bernard2005} show that there are some
uncertainties in this pseudo-critical temperature for the 2+1
flavors with physical quark masses, ranging from $150\:\mathrm{MeV}$
to $190\:\mathrm{MeV}$. We repeat the calculations above with
different values of $T_{0}$, and find that the hierarchy in the
pseudo-critical temperatures also exists for the cases with physical
masses. For example, if we do not rescale the value of $T_{0}$ and
keep it being $270\:\mathrm{MeV}$ as in the pure gauge field case,
we find $T_{c}(P)=212\:\mathrm{MeV}$ and
$T_{c}(l)=242\:\mathrm{MeV}$ for the improved effective potential.
The difference between these two values are obvious. For the
polynomial effective potential this difference becomes smaller but
is also obvious with $T_{c}(P)=227\:\mathrm{MeV}$ and
$T_{c}(l)=240\:\mathrm{MeV}$. One interesting thing is that when the
value of $m_{s}$ is reduced from $140.7\:\mathrm{MeV}$ to
$5.5\:\mathrm{MeV}$ and $T_{0}$ is kept at $270\:\mathrm{MeV}$ as
well, we find that the deconfinement and chiral pseudo-transitions
are almost coincident, with $T_{c}(P)=205\:\mathrm{MeV}$,
$T_{c}(l)=211\:\mathrm{MeV}$ for the improved Polyakov-loop
effective potential $\mathcal{U}_{imp}$ and
$T_{c}(P)=210\:\mathrm{MeV}$, $T_{c}(l)=211\:\mathrm{MeV}$ for the
polynomial effective potential $\mathcal{U}_{pol}$.

\section{Phase Transition in the Case of $\mu\neq 0$ and $T \neq 0$}

In the above sections, we have mentioned that in the presence of
quark chemical potential $\mu$, the (traced) Polyakov-loop $\Phi$
and its conjugation $\Phi^{*}$ satisfying
Eqs.~\eqref{motion_equations} are different from each other. Viz.
\begin{equation}
\Delta\Phi=\frac{\Phi^{*}-\Phi}{2}\label{deltphi}
\end{equation}
develops a finite value. It has also been demonstrated that this
difference originates from the sign problem of the fermion
determinant at finite density, which is unavoidable not only in
Lattice QCD simulations but also in the mean-field
approximation~\cite{Fukushima2006}. However, it was shown that such
a difference between $\Phi$ and $\Phi^{*}$ at finite density may be
not of major qualitative importance in determining the phase diagram
~\cite{Ratti2006a,Sasaki2006}. Therefore, we use the average value
of $\Phi$ and $\Phi^{*}$ to indicate the pseudo-deconfinement phase
transition at finite density, viz.
\begin{equation}
\overline{\Phi} = \frac{\Phi^{*}+\Phi}{2}.
\end{equation}
Solving the four equations in Eqs.~\eqref{motion_equations}, we can
obtain $\Phi$ and $\Phi^{*}$ as functions of temperature $T$ and
quark chemical potential $\mu$, and in turn the $\Delta\Phi$ and
$\overline{\Phi}$ defined above. In the following discussions, we
are only concerned about quantities $\Delta\Phi$ and
$\overline{\Phi}$, therefore, $\overline{\Phi}$ is just denoted by
$\Phi$ from now on.

\begin{figure}[ht]
\centering
\includegraphics[scale=0.9]{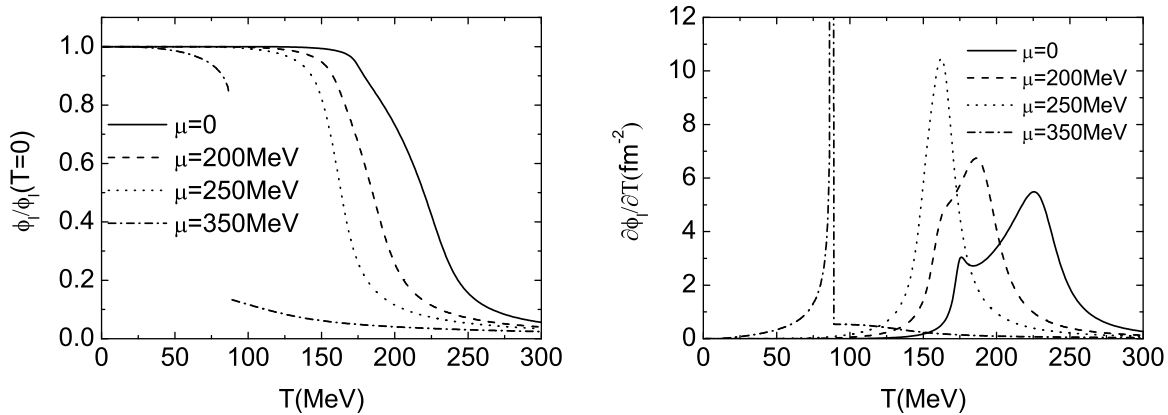}     
\caption{Calculated results of the scaled light quark chiral
condensate (left panel) and its derivative with respect to
temperature (right panel) as functions of temperature for several
values of the quark chemical potential. } \label{f8}
\end{figure}

In this section only the improved Polyakov-loop effective
potential is implemented and $T_{0}$ in this potential is also
chosen to be $215$~MeV. We investigate the cases with physical
quark masses ($m_{l}=5.5\:\mathrm{MeV}$,
$m_{s}=140.7\:\mathrm{MeV}$) and including the flavor-mixing
effect ($K\Lambda^{5}=12.36$), and keep other model parameters
unchanged (except explicit explanations). Fig.~\ref{f8} shows the
calculated temperature dependence of the light quark condensate
and its derivative at several quark chemical potentials. From the
figure, one can recognize easily that, with the increase of
chemical potential, the pseudo-critical temperature of the chiral
transition determined from light quarks decreases and the
crossover becomes more and more narrow and abrupt ( as shown in
the right panel of Fig.~\ref{f8}) and eventually evolves to a
first-order transition at the critical endpoint (CEP)
($T_{CEP}=128\,\mathrm{MeV}$, $\mu_{_{CEP}}=308\,\mathrm{MeV}$).
From the right panel of Fig.~\ref{f8}, one can also notice that
the difference between the chiral pseudo-critical temperature and
that for deconfinement becomes small with the increase of $\mu$,
which means that the chiral phase boundary drops more rapidly than
that of the confinement phase does with the increase of chemical
potential $\mu$. This phenomenon is consistent with the results
shown in Figs.~16 and 17 of Ref.~\cite{Sasaki2006} for the
two-flavor PNJL model.

\begin{figure}[ht]
\centering
\includegraphics[scale=0.9]{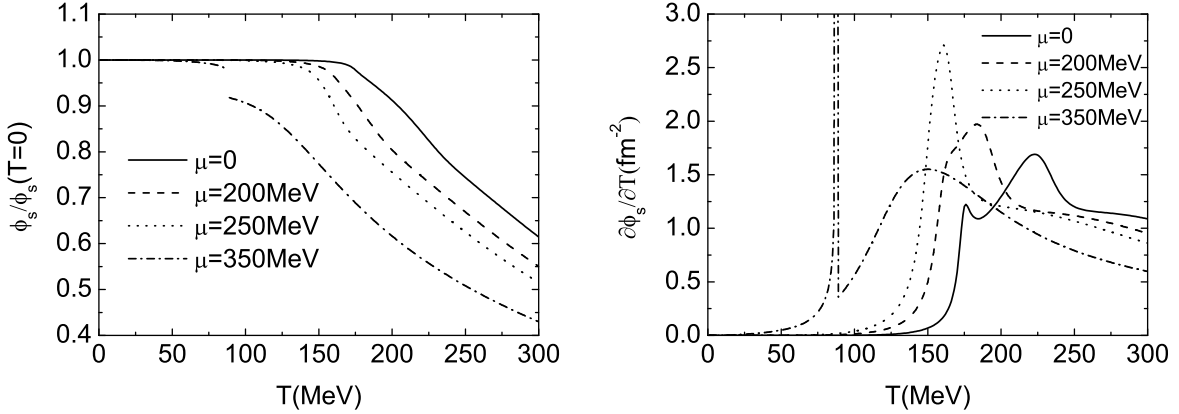}         
\caption{The same as Fig.~\ref{f8}, but for the condensate of
strange quarks.} \label{f9}
\end{figure}

We have also calculated the temperature and chemical potential
dependence of the scaled strange quark condensate and its derivative
$\partial\phi_{s}/\partial T$. The obtained results in the case of
that the chemical potential of strange quarks takes the same values
as for the light quark condensate are illustrated in Fig.~\ref{f9}.
The figure shows evidently that the curves of
$\partial\phi_{s}/\partial T$ in the right panel of Fig.~\ref{f9}
have peaks at the same temperature as the curves of
$\partial\phi_{l}/\partial T$ in the right panel of Fig.~\ref{f8}.
This is reasonable because $\phi_{s}$ is influenced by $\phi_{l}$
through the 't Hooft flavor-mixing interactions. However, higher
temperature is needed to reduce the value of $\phi_{s}$ to approach
zero since the strange quark has much larger current mass.
Therefore, we see that there is a broad peak after the restoration
of the chiral symmetry for light quarks on the curve of
$\partial\phi_{s}/\partial T$ with $\mu=350\:\mathrm{MeV}$ in the
right panel of Fig.~\ref{f9}, which corresponds to the further
reduction of the condensate for strange quarks. It should be noticed
that there are some uncertainties in the location of the broad peak
corresponding to $\mu=350\:\mathrm{MeV}$ because of the high
chemical potential and temperature, since in these regions, the
fermionic distribution function is nonvanishing at momentum beyond
the cutoff of the model, which indicates that those regions are out
of the scope of NJL-like models.

\begin{figure}[ht]
\centering
\includegraphics[scale=0.9]{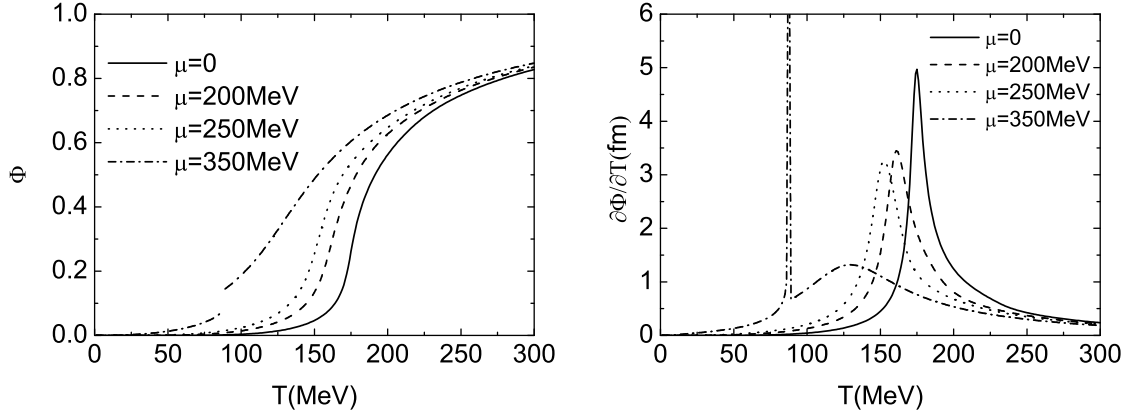}         
\caption{Calculated results of the Polyakov-loop $\Phi$ (left
panel) and $\partial\Phi/\partial T$ (right panel) as functions of
temperature at several values of quark chemical potentials.}
\label{f10}
\end{figure}

\begin{figure}[ht]
\centering
\includegraphics[scale=0.9]{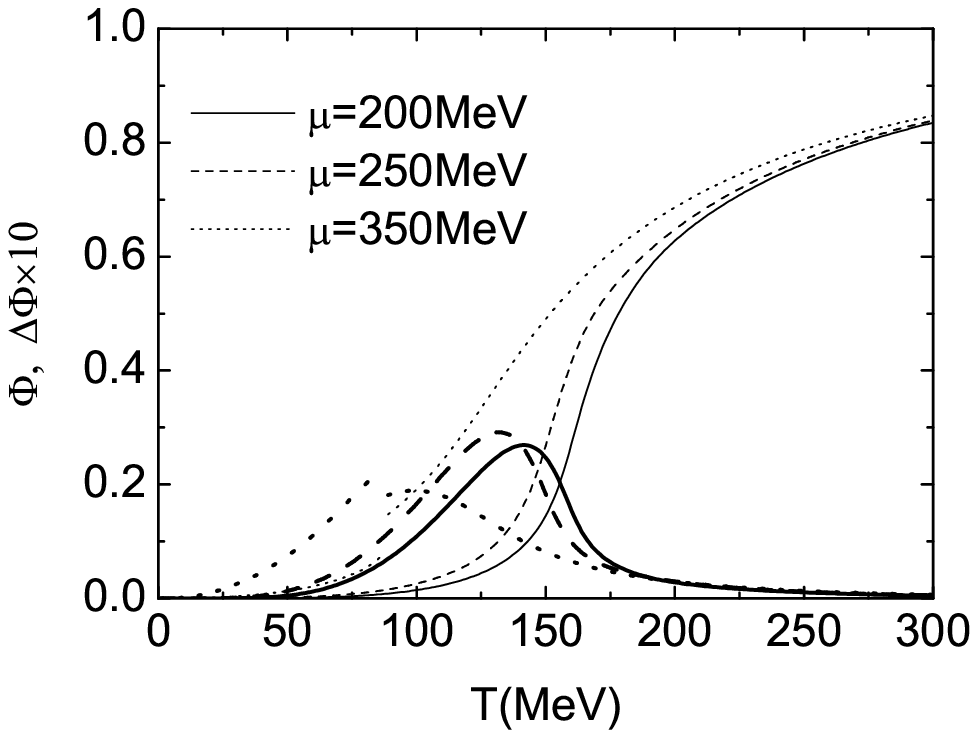}        
\caption{Calculated results of the Polyakov-loop $\Phi$ (thin
curves) and scaled $\Delta\Phi$ (thick curves) as functions of
temperature at several values of the chemical
potential.}\label{f11}
\end{figure}

Fig.~\ref{f10} represents the calculated Polyakov-loop as a
function of temperature at finite quark chemical potential. The
figure manifests evidently that the rapid crossover for the
deconfinement transition at $\mu=0$ is smeared by nonzero $\mu$
and the crossover gets much flatter and milder. This result is
qualitatively consistent with the mean-field result shown in
Fig.~10 of Ref.~\cite{Fukushima2006}. Fig.~\ref{f11} shows the
$\Delta\Phi$ at finite quark chemical potential. We have magnified
the value of $\Delta\Phi$ by ten times in order to compare it to
the $\Phi$ conveniently. As one observes that the value of
$\Delta\Phi$ increases with the temperature below the
deconfinement pseudo-critical temperature $T_{c}(P)$, while above
$T_{c}(P)$, it decreases and approaches to zero.

\begin{figure}[ht]
\centering
\includegraphics[scale=0.9]{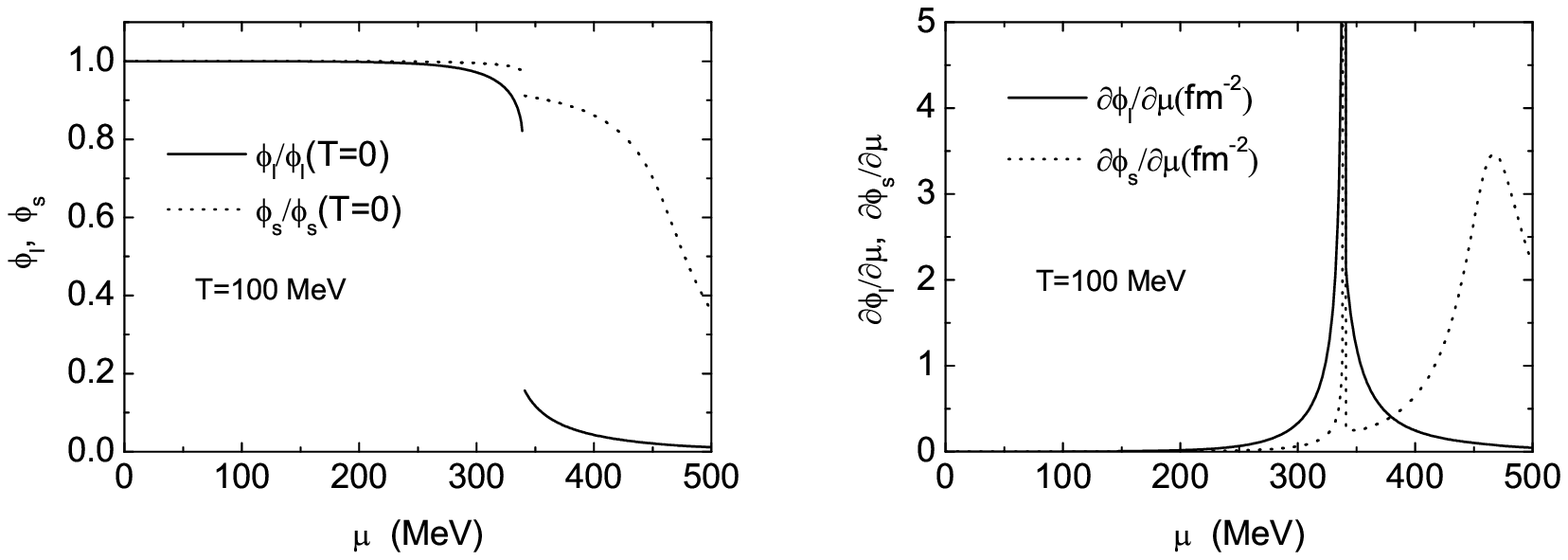}            
\caption{Left panel: calculated quark chemical potential
dependence of the scaled chiral condensates $\phi_{l}$ and
$\phi_{s}$ at temperature $T=100\:\mathrm{MeV}$. Right panel:
plots of $\partial\phi_{l}/\partial\mu$ and
$\partial\phi_{s}/\partial\mu$ with repect to the chemical
potential.}\label{f12}
\end{figure}

\begin{figure}[!htb]
\centering
\includegraphics[scale=0.9,angle=0]{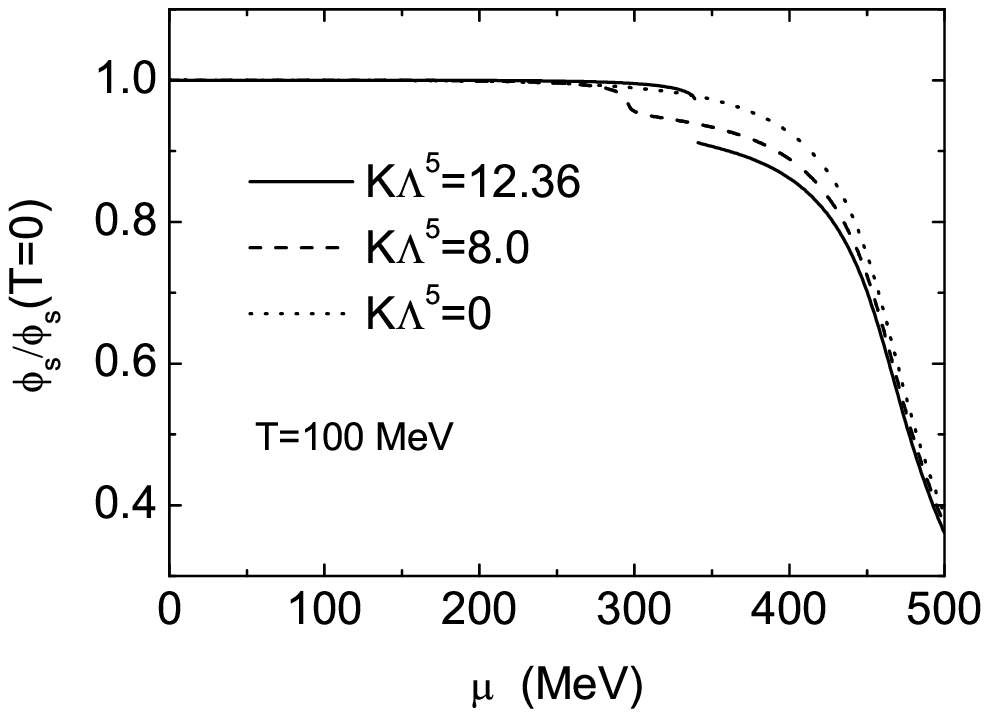}         
\caption{Calculated strange quark condensate as a function of
chemical potential with different values of $K\Lambda^{5}$ at
$T=100\:\mathrm{MeV}$.} \label{f13}
\end{figure}

To discuss the chemical potential dependence of the phase evolution
in detail, we show the calculated quark chemical potential
dependence of chiral condensates $\phi_{l}$ and $\phi_{s}$ at
$T=100\:\mathrm{MeV}$ in Fig.~\ref{f12}. Since the temperature
$T=100\:\mathrm{MeV}$ is below the temperature of CEP
($128\,\mathrm{MeV}$), the chiral phase transition for light quarks
is in first-order, which corresponds to the divergence of the curve
of $\partial\phi_{l}/\partial\mu$ in the right panel of
Fig.~\ref{f12}. Furthermore, besides the influence of the chiral
restoration for light quarks due to the flavor-mixing, exhibiting a
discontinuity in the curves of $\phi_{l}$ and $\phi_{s}$, there is a
smooth crossover for further reduction of the strange quark
condensate at higher values of the quark chemical potential (the
second peak in the curve of $\partial\phi_{s}/\partial\mu$ is
located at about 450~MeV). We also study the influences of the 't
Hooft flavor-mixing strength $K$ on the strange quark chiral
transition with finite quark chemical potential and finite
temperature. The calculated results at temperature
$T=100\:\mathrm{MeV}$ are illustrated in Fig.~\ref{f13}. The figure
shows evidently that, as the flavor-mixing strength becomes weaker,
the temperature of CEP becomes lower. Consequently, the first-order
chiral phase transition of light quarks displaying a discontinuity
in the curve of $\phi_{s}/\phi_{s}(T=0)$ in Fig.~\ref{f13} evolves
to a continuous crossover. While at higher chemical potential for
further reduction of the strange quark chiral condensate, since the
chiral symmetry for light quarks has been restored, the
flavor-mixing interactions that couple different flavors of
condensates can be neglected. Therefore, these three curves in
Fig.~\ref{f13} coincide with each other at large values of the
chemical potential.

\begin{figure}[ht]
\centering
\includegraphics[scale=0.9]{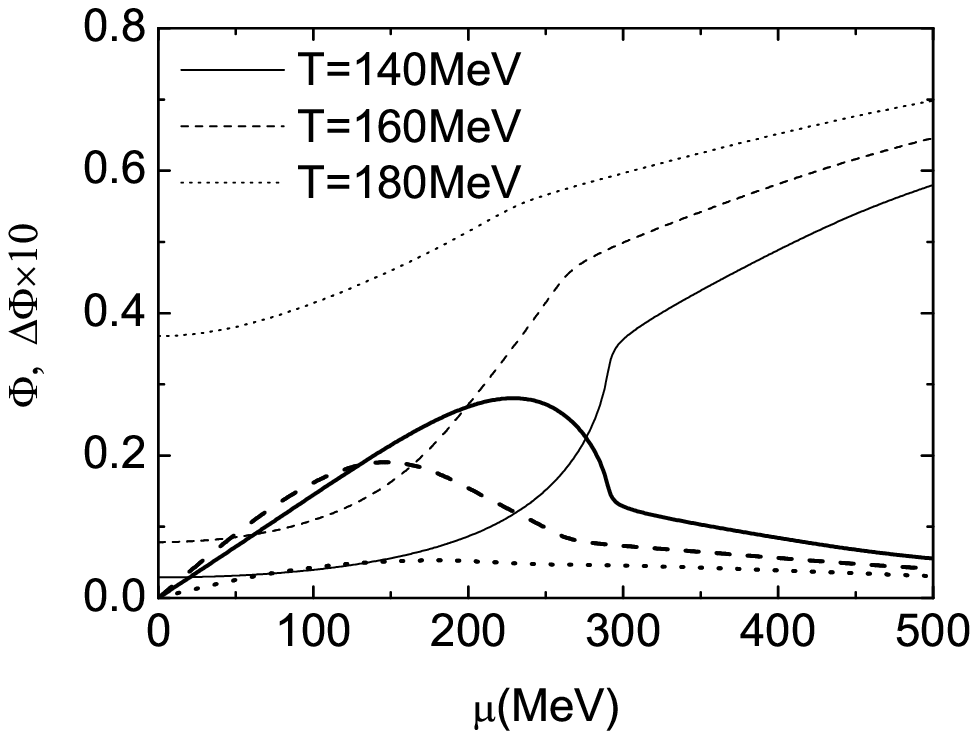}            
\caption{Calculated Polyakov-loop $\Phi$ (thin curves) and the
scaled $\Delta\Phi$ (thick curves) as functions of quark chemical
potential at several values of temperature.} \label{f14}
\end{figure}

We also investigate the quark chemical potential dependence of the
Polyakov-loop $\Phi$ and $\Delta\Phi$. The calculated results at
several temperatures are illustrated in Fig.~\ref{f14}. Such an
investigation is extremely interesting if the chemical potential
is beyond the scope of the current Lattice QCD simulations. It is
easy to infer from the figure that the Polyakov-loop is obviously
suppressed at small $T$ and $\mu$. It indicates a tendency of
confinement. This tendency is reversed with the increase of $T$ or
$\mu$. For example, the value of $\Phi$ corresponding to
$T=180\,\mathrm{MeV}$ (the thin dotted curve) is much larger than
those corresponding to other two values of the temperature at low
$\mu$, since $T=180\,\mathrm{MeV}$ is higher than the
deconfinement pseudo-critical temperature at zero chemical
potential ($T_{c}(P)=175\,\mathrm{MeV}$). The amplified
$\Delta\Phi$ is also shown in Fig.~\ref{f14}, which is much
smaller compared with $\Phi$ in the deconfinement phase.

\begin{figure}[!htb]
\centering
\includegraphics[scale=1.0,angle=0]{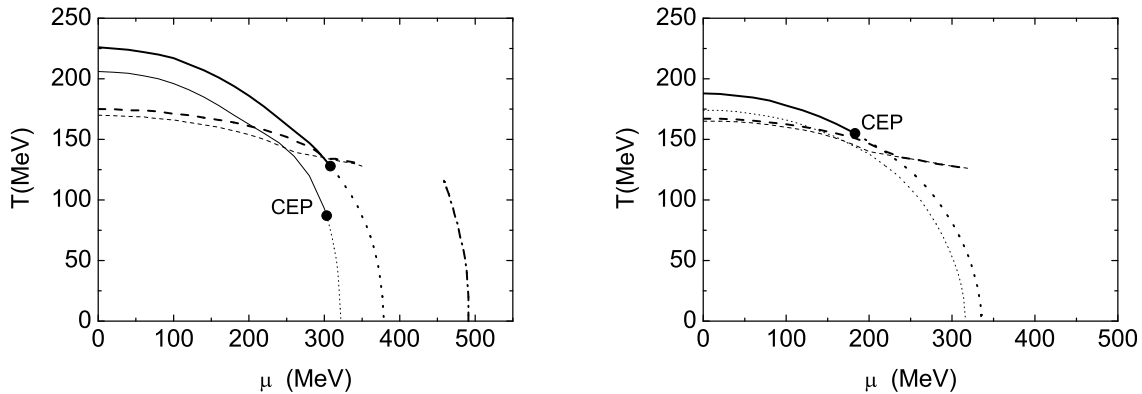}        
\caption{Left panel: calculated phase diagram in terms of
temperature and quark chemical potential of the 2+1 flavor PNJL with
physical masses, viz. $m_{l}=5.5\,\mathrm{MeV}$,
$m_{s}=140.7\,\mathrm{MeV}$, and with different flavor-mixing
interaction strengthes: $K\Lambda^{5}=12.36$ (thicker curves) and
$K\Lambda^{5}=8.0$ (thinner curves). Right panel: phase diagrams for
$m_{l}=m_{s}=5.5\,\mathrm{MeV}$ (thicker curves) and in the chiral
limit $m_{l}=m_{s}=0$ (thinner curves), respectively, with
$K\Lambda^{5}$ fixed at 12.36.} \label{f15}
\end{figure}

As a finality, we present the phase diagrams of the 2+1 flavor PNJL
in Fig.~\ref{f15} and discuss their dependence on the flavor-mixing
interaction strengthes (left panel) and the current masses of quarks
(right panel) for the improved Polyakov-loop effective potential.
The curves for crossover are determined by the location of the peaks
of the derivative of ``quasi" order parameters with respect to $T$
or $\mu$. As for the left panel of Fig.~\ref{f15}, the solid, dotted
curves indicate the crossover, the first-order chiral transition for
light quarks, respectively, separated by the critical endpoint
(CEP). The dashed curves correspond to the crossover for the
deconfinement and we only depict them in the range of
$\mu=0\sim350\,\mathrm{MeV}$. The dash-dotted curves indicate the
``further chiral crossover" for strange quarks and we only depict
them in the temperature range $T \in [0\,, 120]$~MeV, because
NJL-type models are problematic in the range of large $\mu$ and high
$T$ as discussed above. One can observe from the figure that, as the
strength of flavor-mixing interactions becomes weaker, the CEP moves
down to lower temperature. We also find that, when the flavor-mixing
interaction strength $K$ approaches to zero, the CEP moves toward
$\mu$-axis gradually and finally disappears from the phase diagram.
It should be noted that two dash-dotted curves corresponding to two
values of $K\Lambda^{5}$ coincides with each other because of the
reason mentioned above. The same thing occurs for the deconfinement
crossover outside the chiral phase boundary of light quarks.

Furthermore, the dependence of the phase diagram on the current mass
of quarks has been investigated as well. We neglect differences
between the light quarks and the strange quark, and study two cases
with $m_{l}=m_{s}=5.5\,\mathrm{MeV}$, $m_{l}=m_{s}=0$, respectively.
Their corresponding phase diagrams are depicted in the right panel
of Fig.~\ref{f15}. Here, $K\Lambda^{5}=12.36$ is chosen. Since the
current mass of strange quark is the same as those of light quarks,
the strange quark chiral transition coincides with those of light
quarks and there is no dash-dotted line in the phase diagram. when
the current mass of strange quark is reduced from
$140.7\,\mathrm{MeV}$ to $5.5\,\mathrm{MeV}$, both the critical
value of $\mu$ at $T=0$ and that of $T$ at $\mu=0$ decrease, and the
CEP moves toward to the $T$-axis. If the values of both $m_{l}$ and
$m_{s}$ are further reduced, for example in the chiral limit, we
find that the CEP disappears from the phase diagram and there is
only first order chiral transition in the whole range of chemical
potential and temperature, which is illustrated by the thinner
dotted line in the right panel of Fig.~\ref{f15}. This result is
quite consistent with the Pisarski-Wilczek
argument~\cite{Pisarski1984} that the order of the temperature
driven chiral symmetry restoration transition is first order for
three massless quarks. In the framework of the PNJL model, this
could be understood as a result of that the thermodynamical
potential involves a term cubic in the chiral condensate, because of
the 't Hooft interaction. To study the the dependence of the phase
evolution on the flavor-mixing interaction more thoroughly, we
perform a series calculations by reducing the strength of the
flavor-mixing interactions gradually and find that when the
flavor-mixing interaction strength $K\Lambda^{5}$ decreases to a
value of about $5$, the order of the temperature driven chiral
transition is changed from first to second.

\section{Summary and Conclusions}

In summary, we have extended the Polyakov-loop improved NJL model to
the 2+1 flavor case with inclusion of strange quark. This 2+1 flavor
PNJL is a synthesis of the conventional 2+1 flavor NJL model, which
includes the flavor-mixing 't Hooft interaction, and the
Polyakov-loop dynamics governed by a Polyakov-loop effective
potential. Within the framework of such a model, we have studied the
chiral and Polyakov-loop dynamics and their mutual influences to
understand the nature of the QCD phase transitions in the
three-flavor system.

More concretely, we investigate the chiral and deconfinement
crossovers with finite temperature at zero quark chemical potential
with physical current mass of strange quark. Three kinds of
pseudo-critical temperature corresponding to three different
quantities: Polyakov-loop, chiral susceptibility of light quarks and
strange quark number susceptibility, are determined in the PNJL
model. By employing two Polyakov-loop effective potentials existing
in literatures, viz. the polynomial effective potential and the
improved one, we all find that different observables lead to
different values of transition temperature due to the
non-singularity of the crossover, which probably indicate that this
phenomenon is independent of the choice of the Polyakov-loop
effective potential in the PNJL model. However, other effective
potentials need to be developed to verify this result. The hierarchy
in the pseudo-critical temperatures found in our model is consistent
with the recent Lattice QCD results in Ref.~\cite{Aoki2006a}.
However, $T_{c}(\chi_{l})$, $T_{c}(\chi_{s})$ and $T_{c}(P)$ found
in another Lattice simulation~\cite{Cheng2006} almost coincide and
there is not the hierarchy. But we should be careful to use the
results in Ref.~\cite{Cheng2006}, because the lattice spacings used
in Ref.~\cite{Cheng2006} are not in the scaling regime and the
results obtained with their lattice spacings can not give a
consistent continuum limit for $T_{c}$~\cite{Aoki2006a}. Making use
of the two different Polyakov-loop effective potentials, we find
that there is always an inflection point in the curve of strange
quark number susceptibility vs temperature, accompanying the
appearance of the deconfinement phase, independent of the strength
of flavor-mixing interaction, which is also consistent with the
results of Lattice QCD simulations~\cite{Aoki2006a}. Effects of the
current mass of strange quark ($m_{s}$) are studied, too. We find
that the chiral crossover for light quarks moves down to lower
temperature and become more abruptly with the decrease of $m_{s}$,
while the deconfinement crossover is almost not influenced by the
variation of $m_{s}$.

Furthermore, predictions for nonzero quark (baryon) chemical
potential and finite temperature are made in this work. We
investigate the temperature and chemical potential dependence of the
Polyakov-loop $\Phi$, $\Delta\Phi$ and the condensates for light
quarks and strange quarks as well as their mutual interactions. We
find that in the deconfinement phase the value of $\Phi$ approaches
to one and $\Delta\Phi$ is much smaller than that of $\Phi$. We also
give the phase diagram of the strongly interacting matter in terms
of the chemical potential $\mu$ and temperature $T$. It shows that
the critical endpoint (CEP) moves down to the $\mu$-axis and finally
disappears with the decrease of the strength of the 't Hooft
flavor-mixing interaction. On the contrary, the CEP moves toward to
the $T$-axis as the current mass of strange quarks is reduced and
disappears when the chiral limit is approached.

It is well known that the $U_{A}(1)$ symmetry of QCD is explicitly
broken by the axial anomaly coming from the gluon dynamics at
quantum level. It is very interesting to investigate the effects of
the interplay among the $U_{A}(1)$ anomaly, the Polyakov-loop
dynamics and the chiral symmetry breaking at finite temperature and
nonzero chemical potential in the PNJL model. Following
Ref.~\cite{Hansen2007} which studies the properties of pion and
sigma mesons in the two flavor PNJL model, we can analyze the
pseudoscalar mesons and their chiral partners in three flavor case
and their convergence could indicate the restoration of the
$U_{A}(1)$ symmetry with the increase of temperature or chemical
potential. These studies are under progress and we will report it
elsewhere.

\section*{Acknowledgements}

This work was supported by the National Natural Science Foundation
of China under contract Nos. 10425521, 10575004 and 10675007, the
Major State Basic Research Development Program under contract No.
G2007CB815000, the Key Grant Project of Chinese Ministry of
Education (CMOE) under contact No. 305001, and the Research Fund for
the Doctoral Program of Higher Education of China under grant No
20040001010. One of the authors (Y.X. Liu) would also acknowledge
the support of the Foundation for University Key Teacher by the
CMOE. Besides, all the authors acknowledge gratefully the
stimulating discussions with Dr. Lei Chang and Mr. Guo-yun Shao.


\begin{thebibliography}{}
\bibitem{Boyd1996}
G.~Boyd, J.~Engels, F.~Karsch, E.~Laermann, C.~Legeland,
M.~L\"{u}gemeier, and B.~Petersson, Nucl. Phys. {\bf B 469}, 419
(1996).

\bibitem{Engels1999}
J.~Engels, O.~Kaczmarek, F.~Karsch, and E.~Laermann, Nucl. Phys.
{\bf B 558}, 307 (1999).

\bibitem{Fodor2002}
Z.~Fodor, and S.~D. Katz, Phys. Lett. {\bf B 534}, 87 (2002); {\it
ibid}, J. High Energy Phys. {\bf 0203}, 014 (2002).

\bibitem{Fodor2003}
Z.~Fodor, S.~D. Katz, and K.~K. Szabo, Phys. Lett. {\bf B 568}, 73
(2003).

\bibitem{Allton2002}
C.~R. Allton, S. Ejiri, S. J. Hands, O. Kaczmarek, F. Karsch, E.
Laermann, Ch. Schmidt, and L. Scorzato, Phys. Rev. {\bf D 66},
074507 (2002).

\bibitem{Allton2003}
C.~R. Allton, S. Ejiri, S. J. Hands, O. Kaczmarek, F. Karsch, E.
Laermann, and Ch. Schmidt, Phys. Rev. {\bf D 68}, 014507 (2003).

\bibitem{Allton2005}
C.~R. Allton, M. D\"{o}ring, S. Ejiri, S. J. Hands, O. Kaczmarek, F.
Karsch, E. Laermann, and K. Redlich, Phys. Rev. {\bf D 71}, 054508
(2005).

\bibitem{Laermann2003}
E.~Laermann, and O.~Philipsen,
Ann.\ Rev.\ Nucl.\ Part.\ Sci.\  {\bf 53}, 163 (2003).

\bibitem{deForcrand2003}
P.~de Forcrand, and O.~Philipsen,
Nucl.\ Phys.\ {\bf B 642}, 290 (2002); Nucl.\ Phys.\ {\bf B 673},
170 (2003); O. Philipsen, arXiv: hep-lat/0510077; P. de Focrand,
and S. Kratochvila, Nucl. Phys. {\bf B} (Proc. Suppl.) {\bf 153},
62 (2006).

\bibitem{Kratochvila0456}
S. Kratochvila, and P. de Focrand, Nucl. Phys. {\bf B} (Proc.
Suppl.) {\bf 129}, 533 (2004); {\it ibid}, Nucl. Phys. {\bf B}
(Proc. Suppl.) {\bf 140}, 514 (2005); {\it ibid}, Phys. Rev. {\bf
D 73}, 114512 (2006).

\bibitem{delia1}
 M.~D'Elia, and M.~P.~Lombardo,
Phys.\ Rev.\ {\bf D 67}, 014505 (2003).

\bibitem{delia2}
M.~D'Elia, and M.~P.~Lombardo,
 Phys.\ Rev.\ {\bf D 70}, 074509 (2004).

\bibitem{ADGL05} V. Azcoiti, G. Di Carlo, A. Galante, V. Laliena,
Nucl. Phys. {\bf B 723} (2005), 77.

\bibitem{AFHL05} A. Alexandru, M. Faber, I. Hovath, and K.F. Liu,
Phys. Rev. {\bf D 72} (2005), 114513.

\bibitem{Aoki2006a}
Y.~Aoki, Z.~Fodor, S.~D.~Katz, and K.~K.~Szab\'{o}, Phys. Lett.
{\bf B 643}, 46 (2006).

\bibitem{Aoki2006b}
Y.~Aoki, G.~Endrodi, Z.~Fodor, S.~D.~Katz, and K.~K.~Szab\'{o},
Nature {\bf 443}, 675 (2006).


\bibitem{Meisinger9602}
P.~N. Meisinger, and M.~C. Ogilvie, Phys. Lett. {\bf B 379}, 163
(1996); P. N. Meisinger, T. R. Miller, and M. C. Ogilvie,, Phys.
Rev. {\bf D 65}, 034009 (2002).


\bibitem{Pisarski2000}
R.~D. Pisarski, Phys. Rev. {\bf D 62}, 111501 (2000); A. Dumitru and
R.~D. Pisarski, Phys. Lett. {\bf B 504}, 282 (2001), Phys. Lett.
{\bf B 525}, 95 (2002), Phys. Rev. {\bf D 66}, 096003 (2002).

\bibitem{Fukushima2004}
K.~Fukushima, Phys. Lett. {\bf B 591}, 277 (2004).

\bibitem{Ratti2006a}
C.~Ratti, M.~A.~Thaler, and W.~Weise, Phys. Rev. {\bf D 73}, 014019
(2006); C.~Ratti, M.~A.~Thaler, and W.~Weise, nucl-th/0604025.


\bibitem{Ghosh2006}
  S.~K.~Ghosh, T.~K.~Mukherjee, M.~G.~Mustafa, and R.~Ray,
  Phys.\ Rev.\ {\bf D 73}, 114007 (2006).

\bibitem{Ratti2006c}
C.~Ratti, S.~R\"{o}{\ss}ner, M.~A.~Thaler, and W.~Weise, Eur.
Phys. J. {\bf C 49}, 213 (2007) (arXiv: hep-ph/0609218).

\bibitem{Ratti2006b}
S.~R\"{o}{\ss}ner, C.~Ratti, and W.~Weise, Phys. Rev. {\bf D 75},
034007 (2007) (arXiv: hep-ph/0609281).

\bibitem{zhang2006}
Z.~Zhang, and Y.~X.~Liu, Phys. Rev. {\bf C 75}, 064910 (2007)
(arXiv: hep-ph/0610221).

\bibitem{Cheng2006}
M.~Cheng \textit{et al.}, Phys. Rev. {\bf D 74}, 054507 (2006).


\bibitem{Kunihiro1989}
T.~Kunihiro, Phys. Lett. {\bf B 219}, 363 (1989).



\bibitem{Rehberg1996}
P.~Rehberg, S.~P.~Klevansky, and J.~H\"{u}fner, Phys. Rev. {\bf C
53}, 410 (1996).

\bibitem{Karsch2002}
F.~Karsch, Lect. Notes Phys. {\bf 583}, 209 (2002).

\bibitem{Bernard2005}
C.~Bernard \textit{et al.} ({\bf MILC} Collaboration),  Phys. Rev.
{\bf D 71}, 034504 (2005).

\bibitem{Buballa2005}
M.~Buballa,
\newblock Phys. Rept. {\bf 407}, 205-376 (2005).

\bibitem{Rischke2004}
D. H.~Rischke, Prog. Part. Nucl. Phys. {\bf 52}, 197 (2004).


\bibitem{Fukushima2006}
K.~Fukushima, and Y.~Hidaka, Phys. Rev. {\bf D 75}, 036002 (2007)
(arXiv: hep-ph/0610323).



\bibitem{Sasaki2006}
C.~Sasaki, B.~Friman, and K.~Redlich, Phys. Rev. {\bf D 75},
074013 (2007) (arXiv: hep-ph/0611147).


\bibitem{Pisarski1984}
R. D.~Pisarski, and F.~Wilczek, Phys. Rev. {\bf D 29}, 338 (1984).


\bibitem{Hansen2007}
H.~Hansen, W. M.~Alberico, A.~Beraudo, A.~Molinari, M.~Nardi, and
C.~Ratti, Phys. Rev. {\bf D 75}, 065004 (2007).








\end{thebibliography}
\end{document}